\documentclass[preprint,superscriptaddress]{revtex4}
\usepackage{graphics,epsfig}
\usepackage{graphics,epsfig}
\usepackage{epstopdf}
\usepackage{graphicx}
\usepackage{dcolumn}
\usepackage{amsmath}
\newcommand{\be}{\begin{equation}}
\newcommand{\ee}{\end{equation}}
\newcommand{\bea}{\begin{eqnarray}}
\newcommand{\eea}{\end{eqnarray}}

\begin{document}

\title{Thermodynamics and P-v criticality of Bardeen-AdS Black Hole in $4\-D$ Einstein-Gauss-Bonnet Gravity}
\author{Dharm Veer Singh}
\email{veerdsingh@gmail.com}
\affiliation{Department of Physics, Institute of Applied Science and Humanities, G. L. A. University, Mathura, 281406 India.}
\author{Sanjay Siwach}
\email{sksiwach@hotmail.com}
\affiliation{Department of Physics, Institute of  Science, Banaras Hindu University, Varanasi, 221005 India.}

\begin{abstract}
\noindent
We consider the Gauss-Bonnet corrected Bardeen black hole solution in $4\-D$ AdS space-time. The solution is obtained by the limiting procedure adopted by Glavan and Lin in $4\-D$ Einstein-Gauss-Bonnet gravity. The general form of first law of black hole thermodynamics is utilized to calculate various thermodynamics variables. The solution exhibit P-v criticality and belong to the universality class of van-der Waals fluid. The effect of Gauss-Bonnet coupling is investigated on critical parameters and inversion temperature. 
\end{abstract} 
\maketitle

\section{Introduction}
\noindent
Black holes are singular solutions of General Theory of Relativity (GTR) and are completely dark objects according to laws of classical mechanics \cite{rp}. They are characterized by few parameters e.g. mass, charge and angular momentum. Application of laws of quantum mechanics opens up a new window to explore the black hole physics. The black holes seem to  obey the laws of thermodynamics and one can assign temperature and entropy to them. The temperature of the black hole is related with surface gravity and entropy is proportional to the area of event horizon. These laws are tested by their applicability to black hole solutions  in wide variety of theories and seem to be universal. The investigation of microscopic degrees of freedom of some black holes in string theory also lends suppport to the area law of black hole entropy.

The regular black holes \cite{Bardeen:1968,AGB,bron,hayw,zas,lemos} have attracted the attention of physics community recently and the study of these black holes may provide a new window of physics to understand the nature of black hole singularities \cite{Sakharov:1966,Gliner:1966}. Bardeen black hole belong to this class and it has been given the interpretation of magnetically charged black hole \cite{AGB}. Thermodynamics of Bardeen black hole has posed further challenges and Smarr relation seems to be violated \cite{rashid,breton}. If one tries to obtain the correct relation, the area law of black hole entropy is not obeyed and the thermodynamics temperature does not seem to be in agreement with Hawking temperature \cite{sharif}. However, the problem can be rectified if one considers the general form of first law of black hole thermodynamics and include the contribution of various parameters and conjugate potentials carefully \cite{ma14,zhang,gulin,balart,dvs99}.

Black hole thermodynamics can be studied in extended phase space, where Gauss-Bonnet coupling is treated as an independent parameter and the cosmological constant as the pressure of the system \cite{kastor,Cvetic:2010jb,rn,cai}. It has been shown recently that Bardeen-AdS black  hole  belong to the universality class of RN-AdS Black hole and P-v isotherms are similar to a van-der Waals fluid \cite{ag}. The thermodynamics and inversion curves in the Joule-Thompson throttling process are also studied and lend support for the identification of black hole interior with a van-der Waals fluid \cite{Maluf:2018lyu,li19,rizwan,Chen:2020igz}. The Bardeen black hole solutions are also studied in de-Sitter space \cite{singh:17,fr1}. The Gauss-Bonnet generalizations of these solutions in higher dimensions has appeared in \cite{dvs19,kumar19}. 

Recently, a 4-dimensional theory of gravity with Gauss-Bonnet correction is introduced by the re-scaling of Gauss-Bonnet coupling $\alpha\rightarrow\frac{\alpha}{D-4}$ and taking the limit $(D-4)\rightarrow 0$ \cite{gla}. The theory possess all the ingredients of Einstein gravity and circumvent the Lovelock theorem about Gauss-Bonnet correction in $4\-D$ space-time. The spherically symmetric black hole solutions are also obtained in the same paper. The generalization to other black holes has also appeared \cite{pedro,sgg,rahul,zhangw,wen,guo,casalino,konoplya, konoplya2}. Here, we report the regular Bardeen-AdS type black hole solution with Gauss-Bonnet corrections in $4\-D$ space-time.

The paper is organized as follows. We present the Gauss-Bonnet corrected Bardeen-AdS black hole solution in next section. The black hole thermodynamics is introduced in section III. The P-v criticality is discussed in section IV. We outline the summary and directions of future research in the concluding section. 

\section{Bardeen Black Hole solution in $4\-D$ EGB Gravity}
\noindent
Consider the Einstein- Hilbert action coupled to nonlinear electrodynamics in the presence of Einstein-Gauss-Bonnet gravity  with negative cosmological constant;
\be
S =\frac{1}{16\pi }\int d^{4}x\sqrt{-g}\left( R -2\Lambda+\alpha {\cal L_{GB}}-4{\cal L}_m \right),
\label{action}
\ee
where $R$ is the scalar curvature, $\Lambda=-6/l^2$ is the cosmological constant and $\alpha$  is the re-scaled $\alpha\rightarrow\frac{\alpha}{d-4}$ Gauss-Bonnet coupling constant.   ${\cal L}_m$ is the action of nonlinear matter field \cite{AGB}, given by;
\be
{\cal L}_m=\frac{3}{2sg^2}\left[\frac{\sqrt{2g^2F}}{1+\sqrt{2g^2F}} \right]^\frac{5}{2},
\label{matter}
\ee
where $g$ is the magnetic monopole charge, $F=F_{\mu\nu}F^{\mu\nu}$ and $F_{\mu\nu}$ ($\partial _{\mu}A_{\nu}-\partial_{\nu}A_{\mu}$) is the electromagnetic field tensor. The parameter $s$ will be fixed in terms of ADM mass of the black hole as $s=\frac{g}{2M}$. 

We are interested to obtain the static spherically symmetric black hole solution of the theory described by EGB action coupled with non-linear matter Lagrangian. Let us consider the metric ansatz consistent with spherical symmetry; 
\be
ds^2 = -f(r)dt^2+ \frac{1}{f(r)} dr^2 + r^2 d\Omega_{2},
\label{metric}
\ee
where $  d\Omega_{2}$ is the metric of 2-sphere. The ansatz for the field strength is taken as,
\be
F_{\theta\phi}=g \, sin \theta.
\ee
Substituting the general form of metric and ansatz for felectromagnetic field strength into action, we obtain;
\be
S=\frac{1}{2}\int dt\, dr \left(-r(f(r)-1) + \alpha\frac{(f(r)-1)^2}{r}  
+\frac{r^{3}}{l^2}+\frac{g}{s}\left(1-\frac{r^{3}}{(r^{2}+g^{2})^{\frac{3}{2}}}\right)\right)',
\ee
where the prime denotes the derivative with respect to $r$. One can identify the integral of motion as  ADM mass ,  
\be
\frac{1}{2} \left(-r(f(r)-1) + \alpha\frac{(f(r)-1)^2}{r} +\frac{r^{3}}{l^2}+\frac{g}{s}\left(1-\frac{r^{3}}{(r^{2}+g^{2})^{\frac{3}{2}}}\right)\right)=M.
\ee
The above equation is a quadratic equation in $(f(r)-1)$ and can be solved easily to give,  
\be
f(r)=1+\frac{r^2}{2\alpha}\left(1\pm\sqrt{1+4\alpha\left(\frac{2M}{(r^2+g^2)^{3/2}}-\frac{1}{l^2}\right)}\right),
\label{sol}
\ee
where we have used $s=\frac{g}{2M}$.

In the limit $\alpha \rightarrow 0$,  this solution becomes the Bardeen-AdS black hole. The limit $g\rightarrow 0$ corresponds to Gauss-Bonnet AdS-Schwarzschild solution. The horizon of the black hole solution is obtained as the root of the equation $f(r)=0$,
\begin{equation}
1+\frac{r^2}{2\alpha}\left(1\pm\sqrt{1+4\alpha\left(\frac{2M}{(r^2+g^2)^{3/2}}-\frac{1}{l^2}\right)}\right)=0.
\end{equation}
This equation can be solved numerically and for suitable values of the parameters $\alpha$ and $g$ the metric function, $f(r)=0$ admits two solutions $r_{\pm}$. The extremal EGB Bardeen-AdS black holes is obtained if the two horizons becomes degenerate.  Henceforth, we shall treat the outer horizon as event horizon of the black hole. The effect of Gauss-Bonnet coupling on horizon location is shown in the adjoining figures.  The radius of the event horizon decreases with increase in Gauss-Bonnet coupling and the would be black hole may have no horizon at all for the large value of Gauss-Bonnet coupling.
\begin{figure*} [h]
\begin{tabular}{c c c c}
\includegraphics[width=0.55\linewidth]{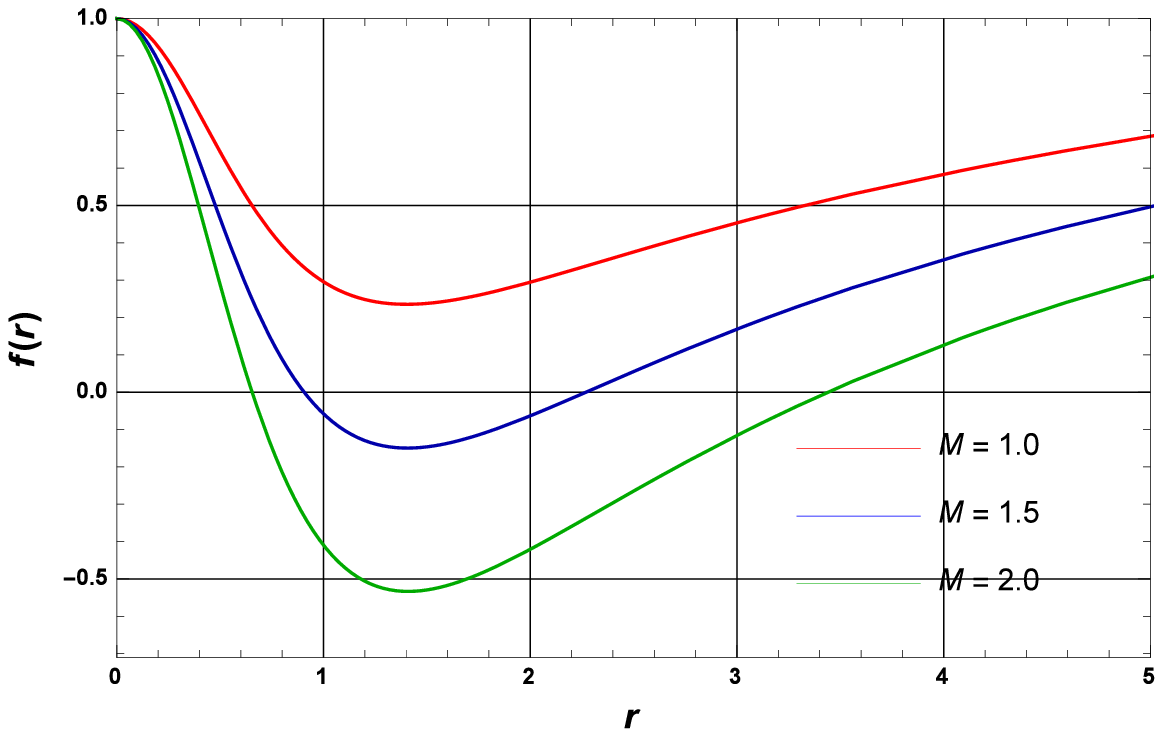}
\includegraphics[width=0.55\linewidth]{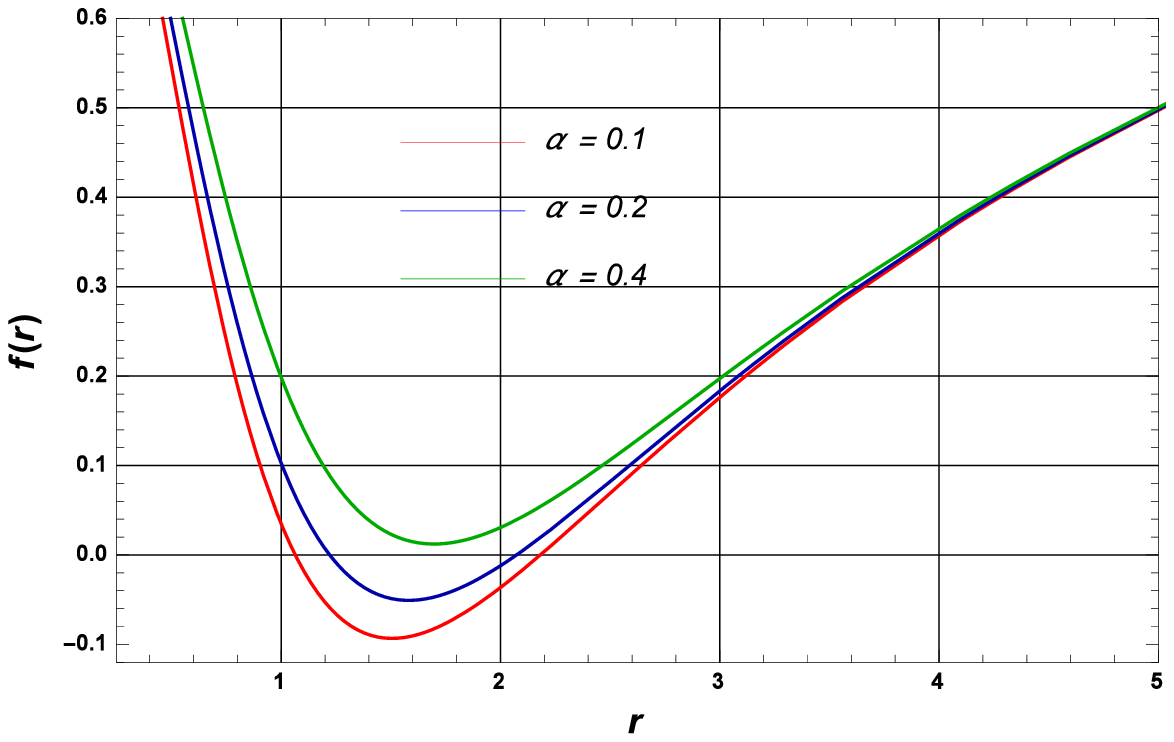}
\end{tabular}
\caption{\label{fig:th} Plot of metric function $f(r)$ vs $r$  for different values of black hole mass for fixed $g=1$ and $\alpha=0$ (left) and for different values of  Gauss-Bonnet coupling for fixed mass $M=1$ and charge $g=1$ (right)   (and $l=20$ in both figures). }
\end{figure*}

The regularity of the black hole solution can be seen by the behavior of the scalar invariants, which are given by,
\bea
\lim_{r\rightarrow0} R & = &\frac{6}{\alpha}\left(\sqrt{1+\frac{8M\alpha}{g^3}-\frac{4\alpha}{l^2}}-1\right),\nonumber\\
\lim_{r\rightarrow0} R_{\mu\nu}R^{\mu\nu}& =&\frac{18}{\alpha^2}+\frac{18}{\alpha^2}\left(\frac{4\alpha M}{g^3}-\frac{2\alpha}{l^2}-\sqrt{1+\frac{8M \alpha}{g^3}-\frac{4\alpha}{l^2}}\right) ,\nonumber\\
\lim_{r\rightarrow0} R_{\mu\nu\rho\sigma}R^{\mu\nu\rho\sigma}  &=&\frac{12}{\alpha^2}+\frac{12}{\alpha^2}\left(\frac{4\alpha M}{g^3}-\frac{2\alpha}{l^2}-\sqrt{1+\frac{8M \alpha}{g^3}-\frac{4\alpha}{l^2}}\right).
\label{inv}
\eea
These invariants have smooth $r\rightarrow 0$ limit and hence the space-time is regular everywhere.

\section{Thermodynamics}
 \noindent
Thermodynamic quantities like energy, temperature, entropy etc. can be assigned to a black hole solution. The mass of the black hole plays the role of enthalpy or heat energy of the black hole.  The solution of the horizon condition $f(r_+)=0$ gives the mass of the black hole in terms of its horizon radius., 
\be
M=\frac{(g^2+r_+^2)^{3/2} }{2l^2 r_+^4}\left(r_+^4+l^2(r_+^2+\alpha)\right).
\label{mass}
\ee
The above expression reduces to the mass of the Bardeen-AdS black hole  in the limit $\alpha \rightarrow 0$ as expected.  

The black hole temperature, also known as Hawking temperature can be defined in terms of surface gravity of the black hole as,  $T_H=\kappa/2\pi$.  The surface gravity can be evaluated using the metric function $f(r)$ and is given by,
\be
\kappa=\frac{1}{2}f'(r)\mid_{r=r_+}.
\ee
Hence, the Hawking temperature for the EGB Bardeen-AdS black hole is obtained as;
\be
T_H=\frac{1}{4\pi r_+}\frac{3r_+^6-2g^2l^2(r_+^2+2\alpha)+r_+^2l^2(r_+^2-\alpha)}{l^2(r_+^2+2\alpha)(r_+^2+g^2)}.
\label{temp}
\ee
The above expression reduces to the temperature of the Bardeen-AdS black hole in the limit $\alpha \rightarrow 0$.  In the adjoining figure we plot the temperature $T$ as a function of horizon radius $r_+$ for different values of Gauss-Bonnet coupling. 
\begin{figure*} [ht]
\begin{tabular}{c c c c}
\includegraphics[width=0.77\linewidth]{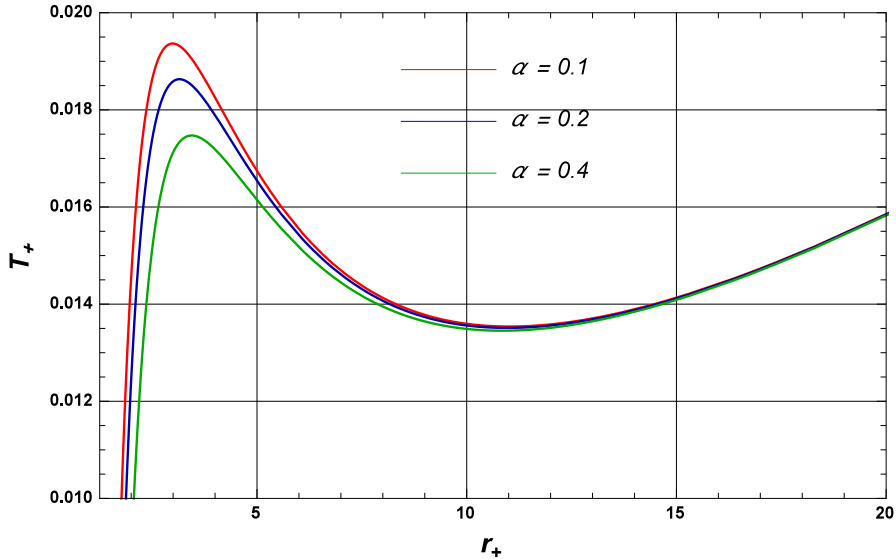}
\end{tabular}
\caption{\label{fig:th1} Plot of temperature $T_+$  vs horizon radius $r$  for different values of Gauss-Bonnet coupling $\alpha$  with fixed value of charge  $g=1$ (and $l=20$) . }
\end{figure*}

The general form of the first law of black hole thermodynamics applied to Bardeen-AdS black hole can be written in the form;
\be
C_M dM=T\,dS + V\,dP + \Phi_q\,dq +\Phi_\alpha \, d\alpha,
\ee
where, $ C_M =\frac{r^3_+}{(r^2_++g^2)^\frac{3}{2}}$ ;  \cite{ma14}, and $P=-\frac{\Lambda}{8\pi}$. $ \Phi_q$ and $\Phi_\alpha$ are the conjugate potentials of the charge and Gauss-Bonnet coupling respectively. The multiplication of left hand side by $C_M$ has to do with the fact the non-linear Lagrangian density of the Bardeen-AdS type black hole depends both on the magnetic charge as well as on the mass of the black hole \cite{zhang,gulin,dvs19}.
\noindent
The thermodynamics variables appearing in the above expression are given by,
\bea
T&=&\frac{3r_+^6-2g^2l^2(r_+^2+2\alpha)+r_+^2l^2(r_+^2-\alpha)}{4\pi l^2r_+(r_+^2+2\alpha)(r_+^2+g^2)}\, ;~~~~~V=\frac{4}{3}\pi r^3_+\, ; \nonumber \\ 
S&=&\pi r^2_+ + 2\pi \alpha \ln r^2_+ \, ; ~~~~~~~~~~~~~~~~~~~~~~~~~~P=\frac{3}{8\pi l^2}\, ; \nonumber \\ 
\Phi_q &=& \frac{3g(l^2(r^2_+ +\alpha) +r^4_+)}{(r_+ l^2(r^2_+ +g^2)}\, ;~~~~~~~~~~~~~~~~~~~\Phi_\alpha = \frac{1}{2r_+} - 2\pi  T \ln r^2_+ .
\label{thermo}
\eea
These variables obey the modified Smarr relation in $4\-D$,
\be
C_M \, M=2 T\, S -2 (P+P_0) \, V + \Phi_q\, q +2\Phi_\alpha \,  \alpha \, .
\ee
Notice the appearance of additional pressure term $P_0 V=2\pi\alpha T$, which is proportional to Gauss-Bonnet coupling and temperature and present even if $P=0$. This term is necessary in order to satisfy the Smarr relation. Thus,  the Gauss-Bonnet corrected system appears to be endowed with an intrinsic pressure. 

In order to analyze the  local stability of the solutions, we consider the heat capacity of the black hole  given by,
\be
C_P=T\left(\frac{\partial{S}}{\partial{T}}\right)_{P,q,\alpha}
\ee
Using the expression of the entropy and temperature as above, the specific heat can be expressed in terms of $r_+$. The expression is too lengthy and we omit it here. In the adjoining figure  we show the plot of $C_P$ as a function of horizon radius $r_+$ for different values of Gauss-Bonnet coupling. 

\begin{figure*} [h]
\begin{tabular}{c c c c}
\includegraphics[width=0.55\linewidth]{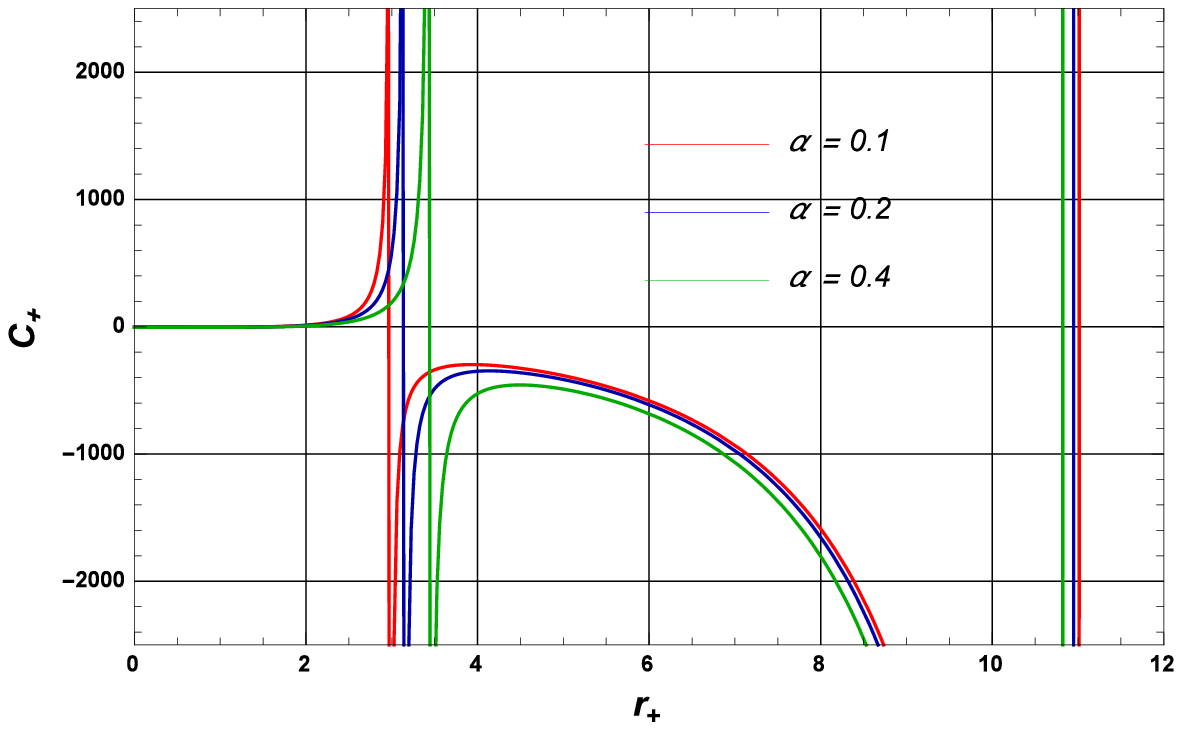}
\includegraphics[width=0.50\linewidth]{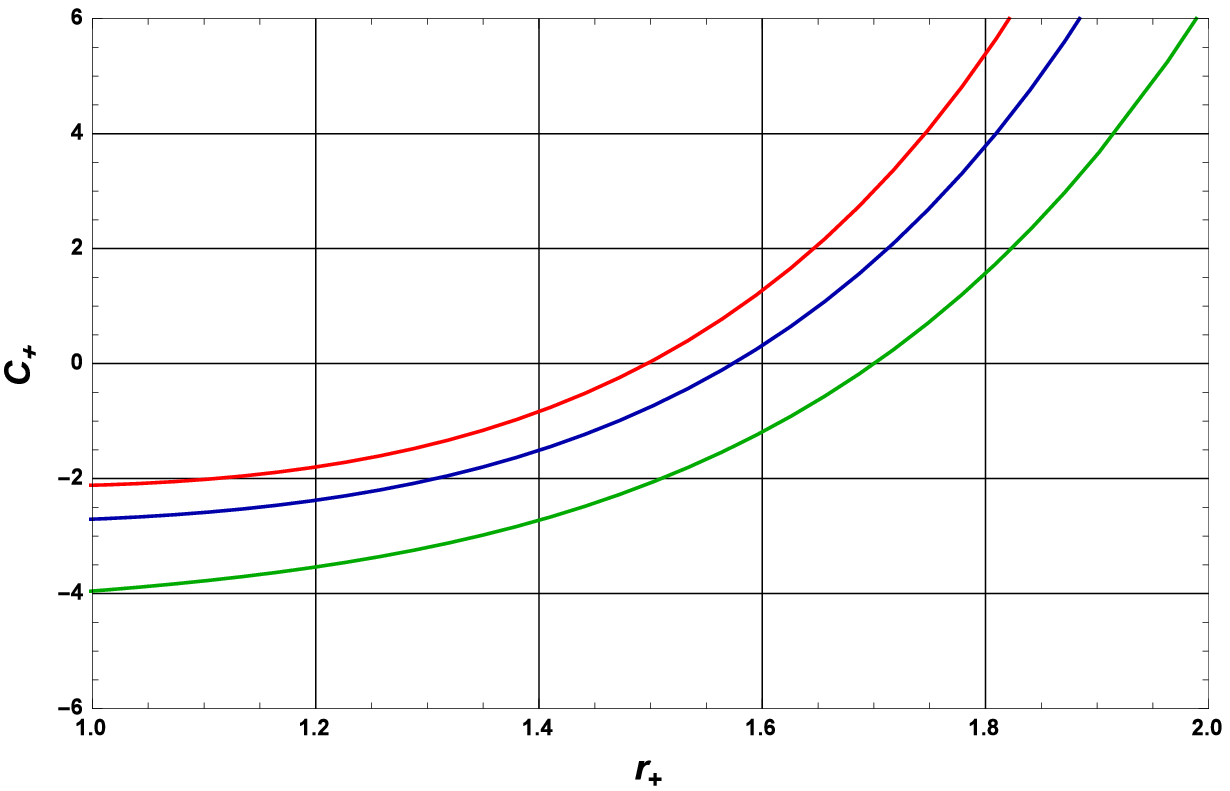}
\end{tabular}
\caption{\label{fig:th2} Plot of specific heat $C_P$  vs horizon radius $r_+$  for different values of Gauss-Bonnet coupling $\alpha$  with fixed value of charge  $g=1$ (left) Insat of the diagram on the left (right)  ( $l=20$). }
\end{figure*}

Let us discuss the stability of the solution for $\alpha=0.1$. Similar considerations apply to other values of Gauss-Bonnet coupling as well.  The heat capacity is negative for $r_+<1.5$ as the temperature becomes negative, which is unphysical and no black hole horizon exist. The upper limit corresponds to extremal  black hole and has zero temperature.  In the range $1.5<r_+<3$ the heat capacity is positive and it corresponds to the stable small black hole solution. The upper limit of $r_+$ corresponds to maxima of temperature. The heat capacity becomes negative beyond this value until we reach the minima of the temperature for $r_+=11$ approx. This region corresponds to unstable black hole solution with intermediate mass. Specific heat becomes positive beyond this value again and this region corresponds to stable large black hole solution.  Thus, the system undergoes a first order phase transition from small black hole to large black hole.

\section{ P-v criticality}

\noindent
The charged AdS black hole (RN-AdS) is known to belong to the universality class of a van-der Walls gas and it undergoes a first order phase transition similar to liquid-gas transition \cite{rn}. The $P-v$ isotherms are also similar and the ratio $\frac{P_c v_c}{T_c}=\frac{3}{8}$, agrees perfectly with the van-der Waals fluid. The Bardeen black hole isotherms are also studied recently and shown to be similar to a van-der Waals fluid \cite{ag}.  The volume used in this case is the specific volume  $v=2 r_+ l^2_P$, where $\l_P$ is the Planck length. We shall put $l^2_P=1$ in all subsequent discussion. 

Consider the equation of state  written in terms of the specific volume,
\be
P=\left( \frac{T}{v}(1+\frac{4g^2}{v^2})+\frac{1}{2\pi v^2}-\frac{4g^2}{\pi v^4}\right)  +\frac{8\alpha}{v^2}\left( \frac{T}{v}(1+\frac{4g^2}{v^2})-\frac{1}{4\pi v^2}-\frac{4g^2}{\pi v^4}\right) .
\ee
Critical constants determined by the conditions,
\bea
\left(\frac{\partial P}{\partial v}\right)_{T}&=&0 \\
\left(\frac{\partial^2 P}{\partial v^2}\right)_{T}&=&0.
\eea
In the absence of Gauss-Bonnet corrections, we get,
\be
P_c=\frac{0.00116}{g^2};~~~~~~~v_c=7.94\,g ~~~~~~~~T_c=\frac{0.025}{g}.
\ee
The ratio,
\be
\frac{P_c\,v_c}{T_c}=0.368,
\ee
is slightly smaller than van-der Waals ratio $0.375$. We have tabulated the values of this ratio for different values of the Gauss-Bonnet coupling below.

The isotherms of EGB Bardeen-AdS black hole are shown in the adjoining figure. We also plot the Gibbs free energy as a function of black hole temperature. Later shows a 'swallow tail'  behavior below the critical pressure, a characteristic of first order phase transition. The  critical temperature is shifted to lower value compared to its value in the absence of Gauss-Bonnet correction.

\begin{figure*} [h]
\begin{tabular}{c c c c}
\includegraphics[width=0.72\linewidth]{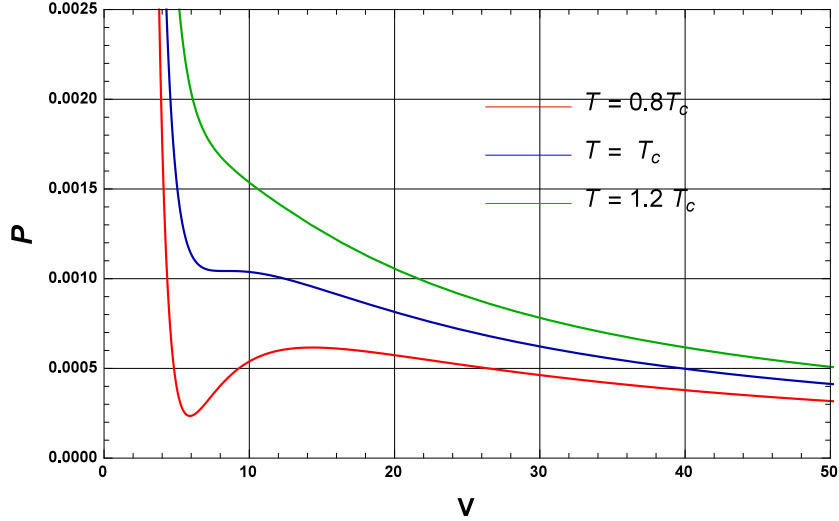}
\end{tabular}
\caption{\label{fig:th3} Plot of pressure $P$  vs volume  $v$  for different values temperature for fixed value of magnetic charge  $g=1$ and Gauss-Bonnet coupling $\alpha=0.1$  . }
\end{figure*}
\begin{figure*} [h]
\begin{tabular}{c c c c}
\includegraphics[width=0.70\linewidth]{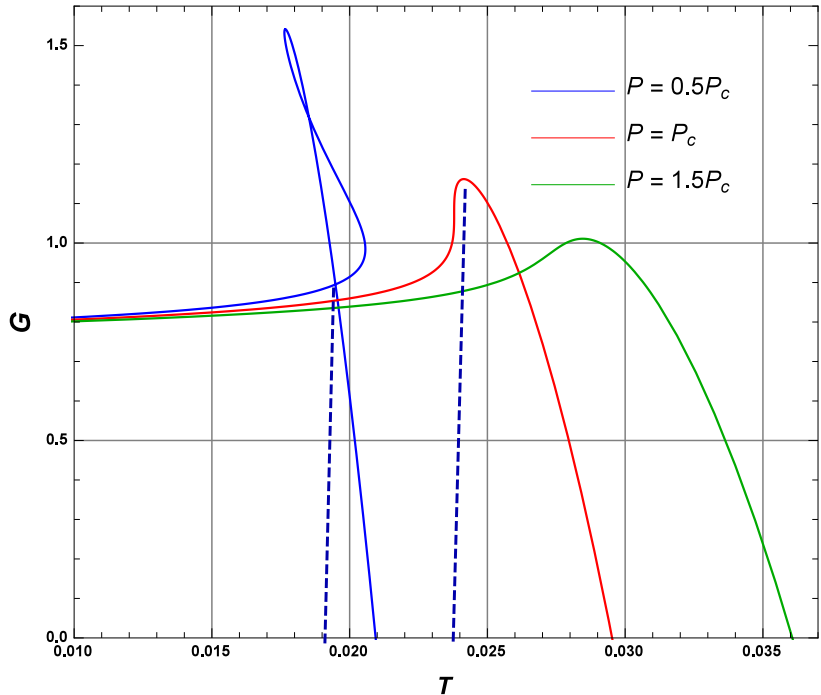}
\end{tabular}
\caption{\label{fig:th4} Plot of Gibbs free energy $G$  vs temperature  $T$  for fixed value of magnetic charge  $g=1$ and Gauss-Bonnet coupling $\alpha=0.1$ . The transition temperature is indicated by dotted line. }
\end{figure*}

\noindent
In the extended phase space, the comparison of the black hole with a van der Walls fluids gives reasonable answers as seen above. Let us consider the Joule-Thomson coefficient  given by,
\be 
\mu=\left(\frac{\partial T}{\partial P}\right)_H=\frac{v}{C_P}\left(\beta T-1\right),
\label{jt}
\ee
where $\beta$ is the coefficient of volume expansion. We can obtain the inversion temperature by setting $\mu=0$,
\be
T_i=v\left(\frac{\partial T}{\partial v}\right).
\ee
The expression can be evaluated and can be expressed in terms of horizon radius $r_+$,
\be
T_i=\frac{2g^2(r^2+2\alpha)^2+r^4A+g^2r^2B}{4\pi r(r^2+2\alpha)^2(r^2+g^2)^2},
\ee
where,
\bea
&&A=8\pi Pr^6+5r^2\alpha+2\alpha^2-r^4(1-48\pi P\alpha),\\
&&B=24\pi P r^6+31r^2\alpha+22\alpha^2+r^2(7+80\pi P\alpha).
\eea

We tabulate the ratio between the minimum inversion temperature and the critical temperature below for different values of the Gauss-Bonnet coupling. These values are less than the value for the van-der Wall fluid, $\frac{T_i}{T_c}=0.75$.

 \begin{center}
\begin{table}[h]
\begin{center}
\begin{tabular}{l|l r l r l| r l r r r}
\hline
\hline
\multicolumn{1}{c|}{ $\alpha$ } &\multicolumn{1}{c}{  } &\multicolumn{1}{c}{ } & \multicolumn{1}{c}{ $P_cv_c/T_c$  }& \multicolumn{1}{c}{}& \multicolumn{1}{c|}{} &\multicolumn{1}{c}{ }&\multicolumn{1}{c}{} &\multicolumn{1}{c}{$T_{i}/T_c$}   & \multicolumn{1}{c}{}& \multicolumn{1}{c}{} \\
\hline
\,\,0\,\, & & &\,\,0.368\,\, &  & & & &\,\,0. 6747\,\,& &
 \\
\,\,0.10\,\,& & & \,\,0.3679\,\,& & & & &\,\, 0.6776\,\,& &
 \\
\,\,0.50\,\,& & & \,\,0.368\,\,&  & & & &\,\, 0.6831\,\,& &
 \\
\,\,1.0\,\,& & & \,\,0.3681\,\,& & & & &\,\, 0.6870\,\,& &
\\
\hline
 \hline
\end{tabular}
\end{center}
\caption{The value of ratio  $P_c v_c/T_c$ and $T_i/T_c$    for different value of Gauss-Bonnet coupling $\alpha$  .}
\label{tab:high}
\end{table}
\end{center}

\section{Summary}
\noindent
In this paper, we have studied the thermodynamics of EGB Bardeen-AdS black in $4\-D$. The general solution is obtained and its horizon structure is analyzed as a function of Gauss-Bonnet coupling. We have followed the procedure of \cite{ma14,dvs99} to obtain correct thermodynamic variables and this is contrary to the claims of entropy and volume modification as reported in the literature. The analysis of heat capacity indicates the first order phase transition from small to large black hole. This is also confirmed by plot of Gibbs free energy as a function of black hole temperature. The P-v criticality is used to obtain the critical constants for different Gauss-Bonnet coupling. The ratio of inversion temperature with critical temperature is also obtained. It would be interesting to explore the $P_c v_c/T_c$ and $T_i/T_c$ ratio further and why do they differ from van-der Waals fluid. The solution can be generalized to rotating black holes.  \\


\noindent
{\bf{Acknowledgments:}} The authors gratefully acknowledge the support of their home institutions.

\end{document}